# Sequential symmetry breakings in charging and discharging Li batteries

Moshe Sheintuch and Olga Nekhamkina

*Departament of Chemical Engineering, Technion, Haifa, Israel 32000.*

The sequential group-by-group charging/discharging in Li batteries with phase-separation thermodynamics was detected by numerical simulations and justified by several experiments published in literature. The present work is the first to quantitatively predict the main characteristics of the sequential symmetry breaking (SB) such as the average Li concentration and the fraction of the low Li part at the SB points. Using a reduced two-zone (low and high concentration) model we derived an approximate solution predicting the effect of initial separation or of imposed parameter perturbation referred to as 'noise' on the SB characteristics. We suggest that the governing parameters that account for the complex effect are the liquid potential drop over the cathode and the noise factor. A bifurcation map showing the lowest current boundary that ensures a homogeneous lithiation/ delithiation as the function of noise is constructed

## I. INTRODUCTION

Symmetry breaking is a phenomenon by which a homogeneous (space-independent) solution is destabilized, leading to a new inhomogeneous stable state. The best-known examples in chemically-reacting systems are the Turing mechanism which is attributed to shaping of tissues in morphogenesis [1], and patterns observed in the Belousov-Zhabotinski (B-Z) reaction with an immobilized catalysts [2]. Following Turing, most of the work in this area was applied to continuously-fed reaction-diffusion systems, in which the activator (the autocatalytic variable) is characterized by small diffusivity, when compared to that of the inhibitor. Under such conditions a steady pattern can emerge with a characteristic length scale (or wave number) that can be predicted from analysis of the model.

Another popular mechanism for pattern formation is that of global coupling, in which a slow-diffusing activator is coupled to a well mixed inhibitor. This mechanism can lead to a stationary front as was demonstrated in catalytic wire systems [3] or to rotating pulse on a catalytic ring [4] or a breathing pulse on a catalytic disk [5]. This mechanism can lead to sustained patterns in electrochemical systems [6]. Studies in electrochemical systems are usually conducted under potentiostatic or galvanostatic (constant current) control, and the latter mode can be satisfied in a system with a well-mixed inhibitor. Patterns under both modes of control, and the transition between them, were observed during anodic dissolution of a nickel wire [7, 8]. In the last decade transient patterns were reported to appear, under galvanostatic conditions, during charging and discharging of batteries, especially of Li-ion batteries which are the type that are the most common today in cell phones, cars, etc. Obviously charging or discharging is a transient process, that cannot reach a sustained pattern, unlike the systems described above. The model employed accounts for an activator ($X_s$ - the relative degree of charge or the dimensionless Li concentration), which is localized (not diffusing), and for several inhibitors. In its simplest form the charging is described by $dX_s / dt = f(X_s)$ where the function $f(X_s)$ reflects information on both thermodynamics and kinetics. It is evident from the model (see below) that under galvanostatic conditions the system can attain a homogeneous, time-dependent, solution and complete charging is achieved at different points (in space) at the same time. Yet the model [9-11] as well as experiments [12] show that when $f(X_s)$ is non-monotonic, and $X_s$ is between the spinodal- (limit-) points, the homogeneous solution undergo symmetry-breaking (SB) leading to a state in which only part of the battery is being charged while, at the same time, the other part is discharged in favor of the first part. When the first part is charged almost to capacity, the second part will start charging but may undergo the same SB process, and so forth and so on.

      The symmetry breaking scenario described above is the main interpretation for many observations of oscillatory behavior during charging and discharging of certain batteries under galvanostatic control. The results are usually described in the plane of the galvanostatic voltage vs the average fractional charge $\bar{X}_s$. To understand this behavior note that the cathode is viewed as a structure made of particles of various sizes. A single particle of phase-separating material is expected to undergo a lithiation and delithiation following the path defined by the potential. In a multi-particle system, the solid solution, when all particles follow the same path, can be maintained as a metastable state even in the spinodal region [13] (large current regimes). However, with small currents the system will undergo the spinodal decomposition: Once a nucleation of a new phase is initiated the following phase separation growth will be completed instantaneously. In a multi-particle system the phase separation process in LiFePO$_4$ was shown to occur as a one-by-one particle (domino cascade [14] or 'mosaic instability' [15]).

It was argued that the particle-by-particle mosaic instability is actually the limiting case of extremely slow charge/discharge rate. Bazant's group simulated the charge/discharge dynamics of nanoparticle phase-separating electrodes [10] and concluded that transition occurs group-by-group one, which is consistent with experimental observations [16-18]. It is argued that since the key prerequisite of the mosaic instability is the free exchange of Li-ions between all electrode particles, the mosaic instability can only be expected when the applied current is sufficiently low [9, 10, 19, 20] For the case of a large current density, there is not enough time for the rapid Li-ion redistribution which is necessary for mosaic instability events. For even higher currents and the consequent high overpotential, the phase separation itself would be suppressed [21-25]. The conducted analysis showed [10] that predictions of the mosaic instability using the Porous Electrode Model (PEM) [26, 27] are in an excellent agreement with the simulations by the particle-level model (in the former case a part-by-part SB is realized).

Here we present a complete picture of the phenomenon using simulations and analysis that predicts the main properties of the group-by group instability in a system subject to noise: the ratio between the Li-rich and -poor phases, the conditions (current, noise, non-uniformities) that escape symmetry-breaking, the cycle intervals in lithiation (or delithiation). We employ the PEM model referred below as the FULL model or using several simplified versions and show that most of the results can be accurately predicted by an approximation that accounts for the two-zone (two-level) distribution of charge.

The structure of this paper is the following: After presentation of the 4-variable one-dimensional mathematical model in Section II, along with one simulation, we suggest several reduced models. In section III we analyze two such reduced models, that ignore gradients in the liquid variables, notably one that accounts for two levels of charging; this reduction allows to derive the main properties of the Symmetry-Breaking, without and with noise. These properties are validated in section IV, and in section V we explain and simulate the effect of the liquid-phase in determining the spatial organization of the pattern.

## II. PROBLEM STATEMENT
### A. Governing equations

At this stage we use the Porous Electrode Model [9, 27] for describing the system: the battery domain is viewed as a continuous media (electrolyte) in which ions diffuse and electrons conduct, and a continuous solid phase media in which the potential $\phi_s$ is space-independent. The charge is stored in the solid phase which is viewed as a cluster of connected particles: the size distribution of the particles and the diffusion resistance between and within them is not addressed here (the size is addressed via a parameter $a_p$, see below). The geometry of the system is as follows: the porous anode is separated from the porous cathode by a separator in which particles are absent and only diffusion takes place in the electrolyte. Li metal foil is the anode for which Li concentration and chemical potential are constant. The kinetic model accounts for the redox reaction: $Li^+ + e^- + FePO_4 \rightleftarrows LiFePO_4$. We employ a 1D model to describe the system behavior within the separator and within the cathode (of the length $L_{sep}$ and $L_{cat}$, respectively) and set $z = 0$ at the anode-electrolyte interface. The liquid phase is also continuous.

We describe several models that are employed in this work: The ***full*** model described below accounts for 4 variables (the concentrations and the potentials in the two (liquid, solid) phases). Later we show that the model can be simplified significantly by ignoring gradients in the liquid phase, which will allow to obtain several analytical results, without significant loss in accuracy.

The *solid phase concentration* follows

$$\frac{\partial C_s}{\partial t} = a_p r_{Li}, \quad z_{sep} < z < z_{tot},  \tag{1}$$

where $a_p$ is the area to the volume ratio of the particles, $C_s = \rho_s X_s$ is the Li ion concentration in the solid phase and $X_s$ - the degree of charge, is the dimensionless concentration with respect to the saturation value ($\rho_s$).

The *reaction rate* is modeled by the modified Butler-Volmer equation

$$r_{Li} = \frac{i_0}{F}\left[\exp\left(-\frac{\kappa F \eta}{RT}\right) - \exp\left(-\frac{(1-\kappa)F\eta}{RT}\right)\right]$$

where $\kappa$ is the transfer coefficient and we set it at the common value of $\kappa = 0.5$ yielding

$$r_{Li} = \frac{2i_0}{F}\sinh\left(-\frac{F\eta}{2RT}\right)  \tag{2}$$

The functions $i_0$ and $\eta$ are the exchange current density and the overpotential respectively. The overpotential is defined as

$$\eta = (\phi_s - \phi_l) - V_{OC} + \frac{\mu_s}{F} \tag{3}$$

where $V_{OC}$ is the plateau value of the open circuit voltage (OCV), $\phi_s$ and $\phi_l$ are the sold and the liquid phase potentials, respectively; the chemical potential $\mu_s$ follows:

$$\mu_s = RT\left[\ln\left(\frac{X_s}{1-X_s}\right) + \Omega(1-2X_s)\right] \tag{4}$$

The function $\mu_s(X_s)$ with $\Omega < 5$ is non-monotonic, exhibiting the local extremums at the spinodal points ($X_{SP1,SP2} = (1 \pm \sqrt{1-2/\Omega})/2$), leading to the instability described below.

The exchange current density defines the rate of charge or discharge, and in the general case it is a function of both the solid and the liquid phase concentrations

$$i_0 = \hat{i}_0 \Psi(C_l, C_s) \tag{5}$$

where $\hat{i}_0$ is the exchange current coefficient. The group-by-group instability was simulated [10, 11] with different types of $\Psi(C_l, X_s)$ including the case $\Psi(C_l, C_s) = 1$ [11]. At this stage we set $\Psi(C_l, C_s) = 1$ to simplify the analysis. In that case, which admits the symmetry $\mu_s(X_s) = \mu_s(1 - X_s)$ (see Eq. 4), the processes of lithiation and delithiation are completely symmetric.

The *liquid phase concentration* is described by

$$\varepsilon \frac{\partial C_l}{\partial t} = \frac{\partial}{\partial z}\left(\varepsilon D_{amb} \frac{\partial C_l}{\partial z}\right) - (1-t_+)a_p r_{Li} \tag{6}$$

Here $\varepsilon$ is the porosity, the diffusion coefficient $D_{amb}$ and the factor $t_+$ are equal:

$$D_{amb} = \frac{D_+ D_- (z_+ - z_-)}{D_+ z_+ - D_- z_-},$$

$$1 - t_+ = \frac{-z_- D_{0-}}{D_+ z_+ - D_- z_-} \tag{7}$$

where $D_\pm$ are the diffusion coefficients of the cation and anion, respectively and $z_\pm$ are the charge numbers.

The *current density* vector follows

$$i = -z_+ v_+ F\left[\frac{F}{RT}(z_+ D_+ - z_- D_-)\varepsilon C_l \frac{\partial \phi_l}{\partial z} + (D_+ - D_-)\varepsilon \frac{\partial C_l}{\partial z}\right] \tag{8}$$

$$-\frac{\partial i}{\partial z} = z_+ F a_p r_{Li} \tag{9}$$

where $v_+$ is the number of cations produced in the reaction. The two latter equations can be combined yielding

$$\frac{\partial}{\partial z}\left[\frac{F}{RT}(z_+ D_+ - z_- D_-)\varepsilon C_l \frac{\partial \phi_l}{\partial z} + (D_+ - D_-)\varepsilon \frac{\partial C_l}{\partial z}\right] = \frac{a_p r_{Li}}{v_+} \tag{10}$$

We focus here on the galvanostatic operation in which the average current density is kept constant

$$i^* = \frac{1}{L_{cat}}\int_0^{L_{cat}} r_{Li,j} F dz = const(t) = 2\hat{i}_0 \sinh\left(-\frac{F\eta^*}{2RT}\right) \tag{11}$$

Here and in the following text we mark the spatial averaged values, which can be time-dependent, by overbar; the superscript star (*) marks the constant values. The applied current is maintained constant by adjusting the value of the electrostatic potential $\phi_s$.

The boundary conditions for system (6), (10) are as follows: no-flux conditions apply for the liquid phase concentration and potential at the cathode-wall interphase ($z = L_{tot}$), at the anode-electrolyte interphase ($z = 0$) a fixed (reference) value $\phi_l$ (=0) is assigned, while the liquid concentration flux is set constant to ensure the constant value of the average salt concentration:

$$z = L_{tot}: \quad \partial \phi_l / \partial z = 0, \quad \partial X_l / \partial z = 0; \tag{12-1}$$

$$z = 0: \quad \phi_l = 0;$$
$$\varepsilon D_{amb}\frac{\partial C_l}{\partial z} = (1-t_+)\int_{L_{sep}}^{L_{tot}} a_p r_{Li,j} dz = const = (1-t_+)\frac{L_{cat}}{F}i^* \tag{12-2}$$

To simplify the following notation we use the dimensionless variables

$$\tilde{\eta} = \frac{\eta}{2RT/F}; \quad \tilde{\mu}_s = \frac{\mu_s}{2RT/F}; \quad \zeta = \frac{z}{L_0}, \quad \tau = \frac{t}{L_0^2/D_{amb}} \tag{13}$$

Assuming that the diffusivities are constants, the dimensionless version of the model takes the form

$$\frac{\partial X_s}{\partial \tau} = a\sinh(-\tilde{\eta}) \tag{14}$$

$$\frac{\partial X_l}{\partial \tau} = \frac{\partial^2 X_l}{\partial \zeta^2} - b\sinh(-\tilde{\eta}) \tag{15}$$

$$\frac{\partial}{\partial \zeta}\left(X_l \frac{\partial \tilde{\phi}_l}{\partial \zeta}\right) = -db\sinh(-\tilde{\eta}) - g\frac{\partial^2 X_l}{\partial \zeta^2} \tag{16}$$

where

$$a = \frac{L_0^2}{\rho_{Li} D_{amb}} \frac{2 a_p \hat{i}_0}{F}; \quad b = \frac{(1-t_+)}{\varepsilon} \frac{\rho_{Li}}{C_l^{initial}} a;$$
$$d = \frac{D_+}{D_+ + D_-}; \quad g = \frac{(D_+ - D_-)}{2(D_+ + D_-)}$$
(17)

The BC (12) take the form:

$$\zeta = 0: \quad \tilde{\phi}_l = 0; \quad \frac{\partial X_l}{\partial \zeta} = -b(\zeta_{tot} - \zeta_{sep})\sinh(-\tilde{\eta})$$
$$\zeta = \zeta_{tot}: \quad \partial \tilde{\phi}_l / \partial \zeta = 0; \quad \partial X_l / \partial \zeta = 0$$
(18)

In the following text the symbol tilde is dropped and we address the dimensionless potentials if otherwise is not stated.

Note, galvanostatic operation yields a constant increase of the average solid phase concentration with time:

$$\frac{\partial \bar{X}_s}{\partial \tau} = a \sinh(-\eta^*)$$
(19)

During homogeneous lithiation the local values $X_s$ follow Eq. (19), the liquid concentration and the potential profiles can be approximated as following (see Appendix A):

$$X_l = \begin{cases} X_l(0) - S\zeta, & \zeta < 1 \\ X_l(0) - s(\zeta_{tot} - 1) + \frac{s}{2}[(\zeta-1)(1+\zeta-2\zeta_{tot})], & 1 < \zeta < \zeta_{tot} \end{cases}$$
(20)

$$\phi_l = \begin{cases} (d-g)\frac{s(\zeta_{tot}-1)}{X_l(0)}\zeta & \zeta < 1 \\ -\frac{(d-g)s}{2X_l(0)}[\zeta^2 + 1 - 2\zeta_{tot}\zeta] & 1 < \zeta < \zeta_{tot} \end{cases}$$
(21)

*Parameters*

We employ the parameters used in Ref. [10]: $D_+ = 1.25 \times 10^{-6} \, cm^2/s$, $D_- = 4 \times 10^{-6} \, cm^2/s$ $\rho_s = 0.0228 \, mol/cm^3$, $C_0 = 1M$, $\hat{i}_0 = 1.75 \times 10^{-6} \, A/cm^2$, $V_{OC} = 3.422V$, $\Omega = 4.5$ $\varepsilon = 0.253$, the particle diameter $d_p = 40$ nm. The SB was detected [10] with the average current values $\bar{i} = i^*/\hat{i}_0$ within the range 0.01-0.2. The corresponding dimensionless $\eta^*$ values are within the range 0.01-0.1 justifying to approximate the function $\sinh(-\eta^*)$ by $(-\eta^*)$ in expressions for the reaction rate (2) and the balance equations (14-16). Note, with assigned $\Omega = 4.5$ the $X_s$ values at the spinodal points ($X_{SP1,SP2} = (1 \pm \sqrt{1-2/\Omega})/2$ ) are equaled to $X_{SP1} = 0.127$, $X_{SP2} = 1 - X_{SP1} = 0.873$.

Because of the symmetry of lithiation/delithiation (the latter coincides with the former under the transformation $X_s \to 1 - X_s$) only the lithiation is considered below and we address the first spinodal point (SP1) as the spinodal point if otherwise is not stated.

## B. Simulation results: the FULL model

A typical solution of a group-by-group instability [10] (also referred to also as part-by-part or domain-by-domain) shows a clearly distinguished stair-case structure with several steps (Fig. 1(a)). Figure 1 illustrates the spatiotemporal $X_s(z,t)$ pattern ($X_s$ is denoted by color, blue for low $X_s$ and yellow for large). In the first interval ($t < t_1$) the solid concentration exhibits a pseudo-homogeneous profiles until $t = t_1$. At that point, a part of the surface of fraction $p = p_1$ adjacent to the wall undergoes, almost instantaneously, discharging while the rest becomes fully charged. Determining the fraction that transformed to the low charged state $p_1 = (z_{tot} - z_1)/(z_{tot} - z_{sep})$, is one of the key points of this paper. The discharged domain (lighter blue) starts again a pseudo-homogeneous charging process which breaks again at $t = t_2$ with a fraction $p = p_2 = (z_{tot} - z_2)/(z_{tot} - z_1)$. This process repeats itself (5 times for the parameters used in Fig. 1) until the whole domain is charged. The temporal evolution is also presented in the ($X_s$ vs $(z,t)$) space (Fig. 1 (b)). Both figures show that the $X_s$ buildup is slow but the SB events are fast.

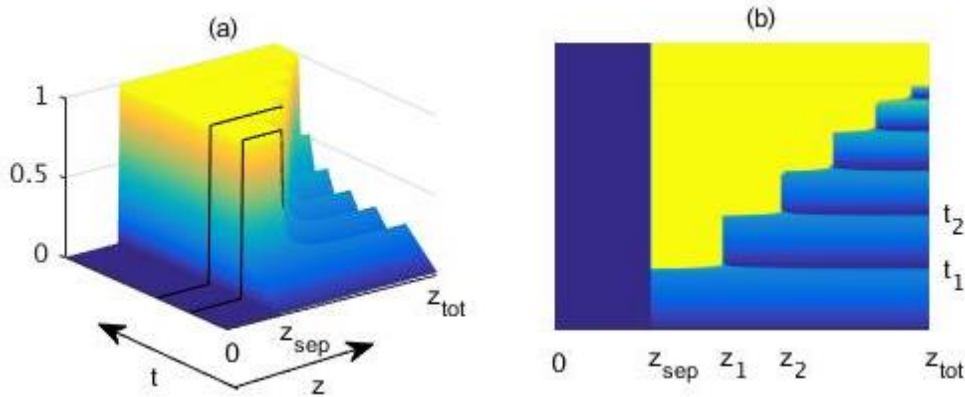

FIG. 1. Typical evolution of the 1D solid concentration ($X_s$) during the lithiation process. Plate (a): 3D presentation in a ($z, t, X_s$) space, lines show several spatial profiles at fixed times; (b) shows a color-scale presentation in a ($z, t$) plane where blue/yellow denote high/low $X_s$. The coordinates ($z_i, t_i$) mark i-th SB. $\bar{i} = 0.02$.

Figure 2 presents the information in terms of spatial $X_s(z)$ profiles at various times around the SB time: these show a jump transition from a low to high values (Fig. 2 (a)) during the short time interval around the transition, $t=t_1$. The liquid potential $\phi_l$ (Fig. 2 (b)) and concentration $C_l$ (Fig. 2 (c)) exhibit smooth profiles that are in a reasonable agreement with approximations of the reaction-diffusion equations (Appendix A) during the pseudo-homogeneous stage. During the short time interval around the transition, $t=t_1$ the $X_s$ in the rich/poor phase may deviate from homogeneity: the spatial profiles can exhibit inverse gradients within the cathode domain (compare solid and dashed lines in Figs 2 (b) and 2 (c)). At the same time the absolute values of $C_l$ and $\phi_l$ variations are relatively small (compare $\phi_l$ with $\phi_s$ presented in Fig. 3), suggesting that their effect can be neglected. However, the role of the liquid phase gradients is crucial in organizing the system as considered in section 4.

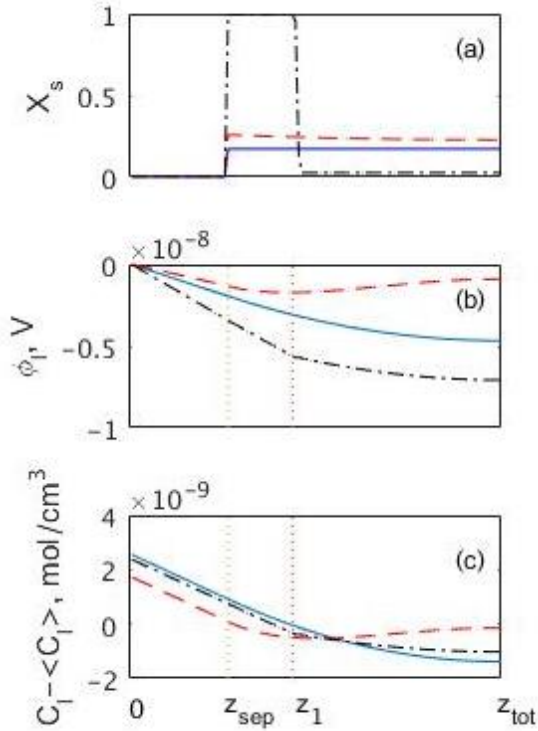

FIG. 2. The evolution of the 1D spatial profiles of three of the four variables of the full model, during the first Symmetry Breaking (SB). Solid, dashed and dashed-dotted lines mark the profiles at times corresponding to the Spinodal Point 1, the start and the end of the first SB, respectively; dotted lines in (b, c) mark the boundary between the separator and the cathode ($z_{sep}$) and location of the first SB ($z_1$). Plate (c) shows the deviation from the average value $C_l$. Parameters as in Fig. 1.

The sequential symmetry breaking process leads to oscillatory behavior of the solid potential ($\phi_s$) when plotted vs the spatially averaged value $\bar{X}_s$ (Fig. 3; recall that $\bar{X}_s$ grows linearly with time), each oscillation corresponds to the lithiation SB cycle described above. Note, for the homogeneous solution the curves $\phi_s$ vs $\bar{X}_s$ are shifted with respect to the equilibrium potential by $\eta^* \simeq 0.5\bar{i}$ (under the approximations above).

The symmetry breaking interpretation of experimental results is based, in most cases, on the oscillatory $\phi_s$ vs $\bar{X}_s$ behavior. Few studies offer direct experimental eveidence: Delman et al [14] characterized the electrochemical deintercalation of LiFePO4 by X-ray diffraction and electron microscopy to show the coexistence of fully intercalated and fully deintercalated individual particles. This result indicates that the growth reaction is considerably faster than its nucleation. The reaction mechanism is described by a 'domino-cascade model'. This was supported by characterization by TEM using Precission Electron Diffraction [29]. Bazant et al [30] capitalized on the distinct colors difference between stages (i.e,.capacity) and employed optical microscopy to give direct information about the lithium concentration in the graphite. Stages were observed to coexist with each other even after extended rest. They concluded that a considerable spatial nonuniformity exists on the microscale. The significance of nonuniformities was emphasized in several theoretical studies (see below).

.

The onset of the instability (Fig. 4) essentially 'overshoots' the spinodal point, where the thermodynamic instability is expected to set in. With increasing current this overshoot increases (is shifted toward the second spinodal point) and with increasing $i^*$ further the charging process becomes homogeneous. Similar results to those in Fig. 3 were obtained in Ref. [10].

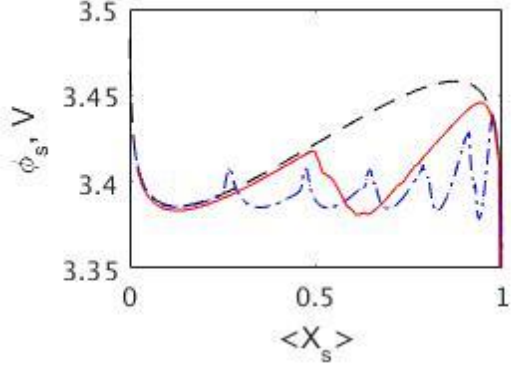

FIG. 3. Typical solid phase potential evolution during the lithiation process with applied currents of $\bar{i}$ =0.02 (dashed-dotted) and $\bar{i}$ =0.10 (solid). Dashed curve shows the equilibrium potential.

The clear two-levels (zones) behavior described above suggests that it will be useful to analyze the dynamics in terms of two populations with rich and poor solid charge levels (the corresponding state variables are marked by superscripts, $X^{\pm}$ and $\eta^{\pm}$; $\phi_s$ is the same within both zones). Since the fraction occupied by the poor-charge level is defined as $p$, we write the overall balance

$$pX^- + (1-p)X^+ = \bar{X}(t) \tag{22}$$

Actually, the surface fraction ($p$) can vary with time. To construct $X^{\pm}$ by the data we assumed that the parameter $p$ changes fast and reaches a certain value that is preserved during the separation cycle.

To complete this picture we draw the trajectories $X^{\pm}$ vs the total $\bar{X}_s$ (Fig. 4; recall that $\bar{X}_s \sim t$). The corresponding model is outlines in Secion 3. We distinguish three sections in the trajectory:

(i) From the initial value to the first spinodal point (SP1) the system exhibits a pseudo-homogeneous solution: the separation in $X_s$ is small: both $X^+$ and $X^-$ are close to the average value $\bar{X}$.

(ii) Beyond SP1 both $X^+$ and $X^-$ increase monotonically and the deviation between the profiles is growing but is still relatively small. Eventually, $X^-$ will change the direction and decrease. Since it is difficult to observe the instability by the increasing separation we choose the maximal $X^-$ value as the Detectable Point (DP, marked by points in Fig. 4) where the separation is distinguished.

(iii) The transition from the DP till the end of the separation cycle is fast: $X^-(t)$ (i.e. $X^-(\bar{X})$) decreases till the minimal value (close to 0), while $X^+(t)$ increases sharply till the maximal value close to 1. The point corresponding to the end of the separation ($\bar{X} = \bar{X}_{EP}$) is marked in Fig. 4 by triangles) and can be addresses as the SB point ($\bar{X}_{SB}$). Note, the peak of the potential $\phi_s$ (Fig. 4 (c)) is close to the DP point.

With increasing time ($\bar{X}$) further the lower branch starts to grow linearly and will undergo another SB etc. The transition from the first and the second cycle shows some overlap where $X^+$ still grows.

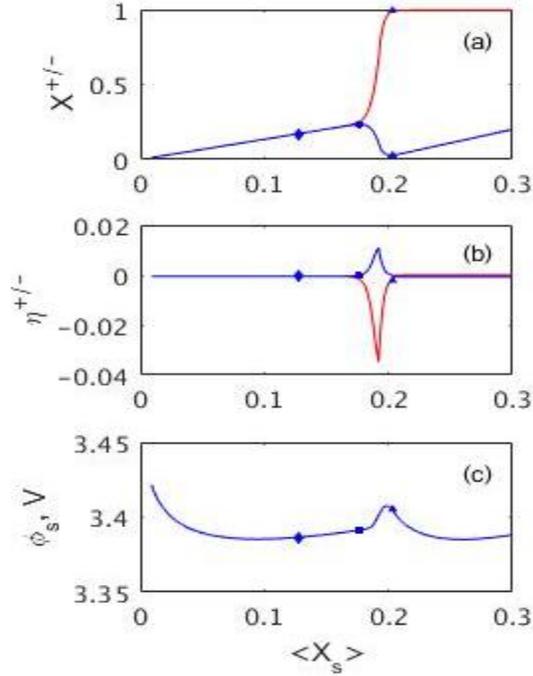

FIG. 4. Evolution of the two-zones solid concentrations ($X^\pm$, a), of the overpotential ($\eta^\pm$, b) and of the solid potential ($\phi_s$, c) vs the total average concentration ($\bar{X}_s$). Symbols (diamonds, points and triangles) mark the first SP, the DP and the point corresponding to the end of the separation, respectively. The parameters as in Fig. 1.

### III. ANALYSIS OF THE APPROXIMATE MODELS
#### A. General statement

The simulation results presented in the previous section show that the deviation of the liquid phase concentration and potential, from their initial value, are of order $10^{-8}$ (with the corresponding units). That justifies to reduce the full model to the *solid* phase *distributed* (SD) model that ignores, for now, the liquid-phase profiles $X_l$ and $\phi_l$, and accounts for $X_s = X_s(t,\xi)$ and $\phi_s = \phi_s(t)$ subject to the galvanostatic-control condition. Simulations conducted with fixed values $\phi_l = 0$; $X_l = 1$ showed that the model exhibits homogeneous lithiation solutions when uniform parameters and homogeneous IC are imposed. But, when perturbations in properties or in IC are accounted for in the SD model, the SD model can exhibit a behavior quite similar to the full four-variable system (14)-(16). This suggests that the noise is the key factor in the SB mechanism; the role of the liquid phase gradients in creating the conditions for SB is addressed in Section IV.

In a real system the lithiation rate will depend on physical properties like activity, geometry (the surface area to the volume ratio of the particles defined by the parameter $a_p$), the site density ($\rho_s$), the diffusivity ($D_{amb}$), so that even with homogeneous initial conditions a separation will be induced by the difference in the growth rate due to nonuniform parameters. The noise factor can be induced in the model either as an instantaneous perturbation (via IC) or as permanent perturbations of the physical parameter (or both). We consider two types of the imposed perturbations: 1) inhomogeneous IC with uniform physical parameters and 2) homogeneous IC with nonuniform parameters.

Realizing that the system usually undergoes symmetry breaking to two levels (zones) of $X_s$ we want to derive an approximate model that predicts these levels and allows to predict the properties of the dynamics. To derive approximations for the SB characteristics the SD model was reduced further to a *solid two-zone* (S2Z) model using the averaged values $X^\pm = X^\pm(t)$ as the state variables. Using this model we obtained approximations for the characteristics of the SB ( $\bar{X}_{DP}$, $p$ ) for both cases. The analysis of case (1) is mathematically simpler and includes the key elements of the proposed approximations, analysis for case (2) follows the former case but does not account for the IC effect. Both models are validated by comparison with numerical simulations of the S2Z model. Simulations of the distributed (SD) model are focused on the nonuniform parameter effect accounting that the SB phenomena was detected by simulations of the full model subject to homogeneous IC.

## B. Approximations

Here we summarize the assumptions and the simplified models used for the following analysis:

i) Gradients in the liquid phase are negligible. For simplicity we set $\phi_l = 0;\ X_l = 1$.

ii) The simplified solid phase distributed (SD) model accounts for two state variables ( $X_s = X_s(t,\xi),\ \phi_s = \phi_s(\tau)$) and is governed by Eq. (11), (14) with $\phi_l = 0;\ X_l = 1$.

iii) The reduced two-zone model (S2Z) accounts for the averaged solid phase variables $X^\pm = X^\pm(t),\ \phi_s = \phi_s(\tau)$.

iv) We assume that the parameter $p$ is preserved during all three stages of the separation.

v) In both models we approximate $\sinh(\eta) \simeq \eta$, which is justified for the range of the parameters used in the paper ($\bar{i} < 0.2$, section II.A)

The S2Z model is governed by the dynamic equations

$$\frac{dX^\pm}{dt} = -a^\pm \eta^\pm; \tag{23}$$

coupled with the galvanostatic condition:

$$pa^-\eta^- + (1-p)a^+\eta^+ = a\eta^* = const \tag{24}$$

and accounts for the overall balance (22). In Eqs. (23) $a^+, a^-$ are parameters that represent a possible non-uniformity of the system characteristics. We examine this case after we study the case of uniform systems.

### 1. Effect of IC separation

To determine the SB characteristics in its simplest (yet meaningful) case we consider the case of the uniform parameter ($a^+ = a^- = a$).

In *Sections (i, ii)* of the trajectory, the divergence between the charges within the both zones (23) follows

$$\frac{d(X^+ - X^-)}{dt} = -a(\mu^+ - \mu^-) \tag{25}$$

Approximating the chemical potential, within each zone ($\mu^\pm$), by its Taylor expansion

$$\mu^\pm = \mu(\bar{X}) + \mu'_{\bar{X}}(X^\pm - \bar{X}) \tag{26}$$

we obtain

$$\frac{d(X^+ - X^-)}{dt} = -a\mu'_{\bar{X}}(X^+ - X^-) \qquad (27)$$

Accounting for $d\bar{X}/dt = -a\eta*$ the latter can be transformed as follows

$$\frac{d(X^+ - X^-)}{d\bar{X}} = \frac{\mu'_{\bar{X}}}{\eta*}(X^+ - X^-) \qquad (28)$$

leading to

$$X^+ - X^- = \delta_0 \exp\left(\frac{\mu(\bar{X}) - \mu_0}{\eta*}\right) \qquad (29)$$

where $\delta_0$ is the initial deviation at $t = 0$ with $\bar{X} = \bar{X}_0$, $\mu_0 = \mu(\bar{X}_0)$.

At the Detectable Point we obtain:

$$X_{DP}^+ - X_{DP}^- = \delta_0 \exp\left(\frac{\mu(\bar{X}_{DP}) - \mu_0}{\eta*}\right) \qquad (30)$$

On the other hand, following the definition of the DP as the point where $X^-$ reaches its maximal value (see Eq. (23): $(dX^-/dt)_{DP} = -a\eta_{DP}^- = 0$) and the condition of a constant average current (24) we obtain

$$\eta_{DP}^- = \phi_{sDP} - V_{OC} + \mu_{DP}^- = 0;$$
$$\eta_{DP}^+ = \phi_{sDP} - V_{OC} + \mu_{DP}^+ = \eta*/(1-p); \qquad (31)$$

yielding

$$\mu_{DP}^+ - \mu_{DP}^- = \frac{\eta*}{1-p} \qquad (32)$$

Accounting for approximation (26) we have

$$\left(\mu'_{\bar{X}}\right)_{DP}(X^\pm - \bar{X})_{DP} = \frac{\eta*}{1-p} \qquad (33)$$

Combining Eqs. (30) and (33) we obtain

$$\delta_0 \left(\bar{\mu}'_{\bar{X}}\right)_{DP} \exp\left(\frac{\mu(\bar{X}_{DP}) - \mu_0}{\eta*}\right) = \frac{\eta*}{(1-p)} \qquad (34)$$

To use this equation we complement the relation above with another mass balance relation between $\bar{X}_{DP}$ and $p$: The end of the separation process can be defined as the point where $X^-$ reaches the minimal value. However, the detailed analysis of the dynamics within this stage is rather cumbersome as a linear approximations similar to (26) cannot be applied. For simplicity we assume that stage (iii) is fast (actually, instantaneous), so the average concentration at the DP and at the end of the separation cycle coincide yielding a linear relation between $\bar{X}_{DP}$ and $p$:

$$\bar{X}_{DP} = \bar{X}_{EP} = p(X^-)_{EP} + (1-p)(X^+)_{EP} \tag{35}$$

Assuming also that at the end of the separation $(X^-)_{EP} \simeq 0$, $(X^+)_{EP} = 1$ we obtain

$$(1-p) = \bar{X}_{DP} \tag{36}$$

which provides an estimate for $p$.

For a given current ($\eta^*$), algebraic system (34), (35) (or (36)) with the assumption of an instantaneous section (iii) allows to obtain $\bar{X}_{DP}$ and $p$ as the functions of the initial separation ($\delta_0$). The low and the high concentrations at the DP ($X^{\pm}_{DP}$) can be defined separately using Eq. (22) with the known $\bar{X}_{DP}$, $p$ and $\delta_0$ values. Obviously, the deviations $\Delta X^{\pm} = X^{\pm} - \bar{X}$ admit the following relation:

$$\left(\frac{\Delta X^+}{\Delta X^-}\right)_{DP} = -\frac{p}{1-p} \tag{37}$$

Note, Eq. (36) allows to predict the limiting values of the separation fraction $p$, using for the Detectable Point the extreme SP values:

$$\lim_{\bar{X}_{DP} \to \bar{X}_{SP2}} p = \bar{X}_{SP1} \ ; \quad \lim_{\bar{X}_{DP} \to \bar{X}_{SP1}} p = \bar{X}_{SP2} \tag{38}$$

For the set of parameters used in the paper $\bar{X}_{SP1} = 0.127$ and so $p$ varies within the range [0.127 0.873].

Recall, the imposed IC perturbations decay within section (i) along a stable branch $\mu(\bar{X})$, as follows from Eq. (29), and are highly dependent on the IC value $\bar{X}_{IC}$ (i.e. the distance from the spinodal point). Therefore, to analyze the initial separation effect it is reasonable exclude section (i) from the consideration and to impose the perturbations at the SP ($\delta_{SP}$). Approximation (34) while accounting for (36) can be presented as following

$$\delta_{SP}(\bar{\mu}'_{\bar{X}})_{DP} \exp\left(\frac{\mu(\bar{X}_{DP}) - \mu_{SP}}{\eta^*}\right) = \frac{\eta^*}{\bar{X}_{DP}} \tag{39}$$

## 2. Nonuniformity effect.

As discussed above, the lithiation rate will depend on physical properties like activity, geometry, so that even with homogeneous initial conditions ($X^-(0) = X^+(0)$) a separation will be induced by the difference in the growth rate due to nonuniform parameter. Let the deviation of any physical property (e.g., geometrical parameter), within both zones, be accounted via a perturbation of the coefficient $a$, by a small parameter ($\alpha > 0$) weighted with $p$ to ensure that the average current balance is maintained. Then dynamic equations (23) can be presented as following

$$\frac{dX^-}{d\tau} = -a\left(1 - \frac{\alpha}{p}\right)\eta^-;$$
$$\frac{dX^+}{d\tau} = -a\left(1 + \frac{\alpha}{1-p}\right)\eta^+; \tag{40}$$

We pursue the analysis of Eqs (40), which can be constructed analytically, and for simplicity we assume equal IC for both zones: $\tau = 0$: $X^+ = X^- = X_{IC} \ll X_{SP1}$.

Following the approach proposed for the uniform property case we obtain for sections (i) and (ii) of the trajectory (see relation (B14) Appendix B for details):

$$\eta^* = \frac{\alpha}{p}\left(1 + \frac{\alpha}{1-p}\right) J(\bar{X}_{DP})\mu'(\bar{X}_{DP})\exp\left(\frac{\mu(\bar{X}_{DP})}{\eta^*}\right) \tag{41}$$

where the function $J(\bar{X})$:

$$J(\bar{X}) = \int_{\bar{X}_{IC}}^{\bar{X}} \exp\left(-\frac{\mu(\bar{X})}{\eta^*}\right) d\bar{X} \tag{42}$$

can be tabulated a'priory. Note, Eq. (41) presents an algebraic relation with respect to $\bar{X}_{DP}$.

As in the case of uniform properties, section (iii) of the trajectory is assumed to be fast and relation (35) is preserved. In the limiting case $p = 1 - \bar{X}_{DP}$ Eq. (41) is reduced to the following form

$$\eta^* = \frac{\alpha}{1 - \bar{X}_{DP}}\left(1 + \frac{\alpha}{\bar{X}_{DP}}\right) J(\bar{X}_{DP})\mu'(\bar{X}_{DP})\exp\left(\frac{\mu(\bar{X}_{DP})}{\eta^*}\right) \tag{43}$$

For a given current ($\eta^*$) algebraic system (43) and (36) allows to obtain $\bar{X}_{DP}$ and $p$ as functions of the imposed permanent perturbation. Highly non-uniform systems are characterized by a Detectable Point that occurs close to SP1, while improving the uniformity of the system will push $\bar{X}_{DP}$ toward SP2 or even avoid SB bifurcation altogether.

*Sequential SB:* Eq. (43) predicts the properties of the first SB. To predict the sequential SB bifurcations we assume that within each cycle the following lithiation process takes place: the highly-charged zone ($X^+$) is inactive, while the zone occupied by the low solid concentration ($X^-$) repeats the previous cycle under an increased (due to previous breaks) current, i.e. the Li concentration is gradually accumulated till the appropriate $\bar{X}_{DP}$ and eventually the next separation occurs. We assume that relation (43) is valid for the subsequent SB events. Then, we consequently obtain: the first separation takes place with $\eta_1 = \eta^*, p = p_1$, $(X_{SB})_1 = X_{DP} = 1 - p_1$; the second cycle occurs under the current $\eta_2 = \eta^*/p_1$ yielding the local value $p = p_2$, while the total fraction occupied by the low solid concentration is equal to the product $p_1 p_2$ and the total average concentration is equal $(\bar{X}_{SB})_2 = 1 - p_1 p_2$. Following such an approach, the k-th cycle takes place under a current of $\eta_k = \eta^*/\prod_{m=1}^{k-1} p_m$ yielding the local value $p = p_k$, while the total low Li domain and the average concentration are equal

$$p = \prod_{m=1}^{k} p_m$$
$$(\bar{X}_{SB})_k = 1 - \prod_{m=1}^{k} p_m \qquad (44)$$

The period of oscillations with $\bar{X}$ (can be translated to the frequency with time) is defined by the difference in the following two cycles:

$$(\bar{X}_{SB})_k - (\bar{X}_{SB})_{k-1} = (1 - p_k)\prod_{m=1}^{k-1} p_m \qquad (45)$$

Note, that while the 'local' DP point $X_{DP}$ belongs to the spinodal interval ($\bar{X}_{SP1} < \bar{X}_{DP} < \bar{X}_{DP2}$), the values of $(\bar{X}_{SB})_k$ can cover the whole domain $\bar{X}_{SP1} < \bar{X}_{SB} < 1$.

## C. Validation

Validation of the proposed approximations is conducted by comparing their prediction with those of the direct simulations of either the Solid-2-Zone (S2Z) or the Solid Distributed (SD) models. In the S2Z model we use an approximate value of the parameter $p$. Such simulations allow to plot $\bar{X}_{DP}$ as the function of either $\delta_0$ (i.e., separation with uniform parameters) or as a function of the nonuniformity parameter (with homogeneous IC) and to verify the assumptions made regarding the system behavior. Simulations of the SD model are more informative as they allow to obtain the phase-separation fraction ($p$) as well as $\bar{X}_{DP}$. The validation using SD model was conducted only for the case of the nonuniform parameters with homogeneous IC (resemble the SB phenomena detected by simulations of the full model subject to homogeneous IC). Also we focus on the parameters corresponding the first SB. The analysis of the sequential SB for the nonhomogeneous IC requires to define the IC at the beginning of the next cycle and is not addressed, while we used fixed values at the end point $X^- = 0$, $X^+ = 1$ to obtain approximation (36). The sequential SB for the nonuniform parameter case is discussed at the end of the section.

### *1. Effect of IC separation.*

The SB properties for a system with uniform parameters are determined by the initial separation $\tau = 0: \bar{X} = \bar{X}_{IC}, \delta = \delta_0 > 0$ ; this was analyzed in section B.1 resulting in Eq. (34), which was transformed to relation (39) in terms of $\delta_{SP}$. In section B.2 we analyzed the outcome of a system with $\delta_0 = 0$ but with non-uniformity and that led to Eq. (43) that expresses the relation between $\bar{X}_{DP}, \alpha$ and the current ($\eta*$). That enables us to compare the two effects of inhomogeneity of IC or of the parameters.

Actually both effects are determined by the separation at SP ($\delta_{SP}$). Approximate solution of a two-zone (S2Z) model subject to imposed *permanent parameter perturbations* yields (Appendix B, Eq. (B9)) the following separation at the spinodal point

$$\delta_{SP} = \frac{\alpha}{p(1-p)} J(\bar{X}_{SP}) \exp\left(\frac{\mu_{SP}}{\eta*}\right) \tag{46}$$

Figure 5 presents the $\delta_{SP}(\alpha)$ dependence expressed above, implying that points on line $\delta_{SP}(\alpha)$ will have similar characteristics when assuming non-uniform IC or non-uniform parameters. With increasing $\bar{i}$ the system is more sensitive to $\alpha$, but is still within the feasible domain of parameters showing that the assumption on small deviation of $X^+$ and $X^-$ from the average value used above (section B.1) are reasonable.

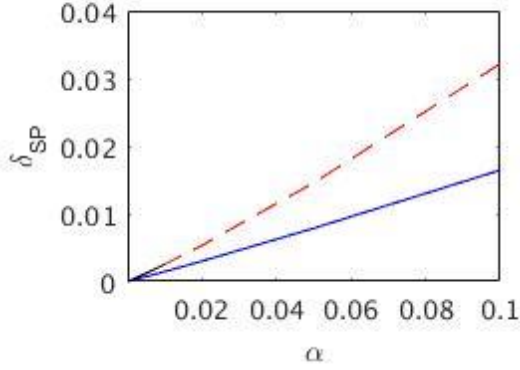

FIG. 5. The effect of the non-uniformity parameter ($\alpha$) on the separation between the high/low concentration at the spinodal point ($\delta_{SP}$) following approximation (46). $\bar{i} = 0.02$ (solid) and 0.2 (dashed).

Following the comments above, the simulations of the S2Z model were conducted within sections (ii) and (iii) of the trajectory with IC $\tau = 0 : \bar{X} = \bar{X}_{DP}$ and an imposed initial perturbation $\delta_{SP}$ which is distributed between $X^-(0)$ and $X^+(0)$ using fixed $p$ value obtained by approximations (section B.1). The initially quasi-homogeneous solution is splitted (Fig. 6 similar to the full system simulations, Fig. 4) into two branches (the simulations were stopped at the end of the first SB defined as the minimum $X^-(\bar{X})$ marked by triangles in Fig. 6). With relatively large $\delta_{SP}$ ($10^{-2}$ with $\bar{i} = 0.02$) the Detectable Point occurs close to the SP1 and it is shifted towards the second SP with decreasing $\delta_{SP}$, for sufficiently small $\delta_{SP}$ no SB is observed. Increasing current has a similar effect, the SB is delayed: the DP is shifted towards the second SP and eventually leaves the spinodal domain leading to form a homogeneous lithiation process. The effect of increasing current was predicted in literature [9, 10, 12, 19, 20]. Note, with increasing $\bar{i}$ the duration of the last section (the transition from the DP till the end of the separation) is prolonged, while the value of $X^-$ at the end of the separation gradually increases. Recall that the derivation of the approximations assumed instantaneous transition from DP to the end of the cycle.

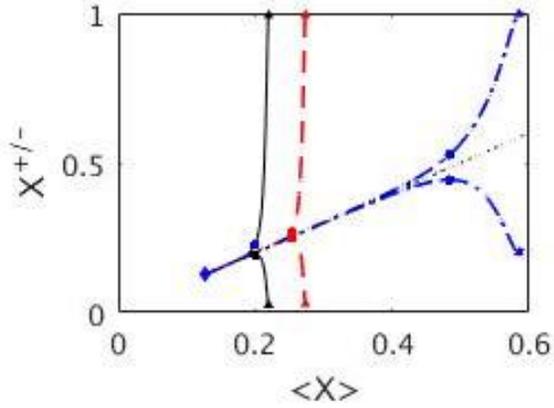

FIG. 6. The effect of the imposed perturbation at the spinodal point ($\delta_{SP}$) and of the current ($\bar{i}$) on the evolution of $X^+$ and $X^-$ simulated with the S2Z model (Eq. (23)-(24)). The sets ($\bar{i}, \delta_{SP}$) are (0.02, $10^{-4}$ - solid), (0.02, $10^{-8}$ - dashed), (0.2, $10^{-4}$ - dashed-dotted). Symbols (diamond, points and triangles) mark the characteristic points as in Fig. 4.

The agreement between the simulated and the approximated results ($X^+$ and $X^-$ are shown separately in Fig. 7) is excellent to very good with large or medium currents (large values of $\bar{X}_{DP}$, Fig. 7 (b)) and as the current decreases $\bar{X}_{DP}$ tends to the first spinodal point (Fig. 7 (a)). These results validate the approximations proposed in section B.1 which are the key elements of the analysis. The DP's are shifted to higher $\bar{X}$ values as the imposed perturbation ($\delta_{SP}$) declines (Fig. 7).

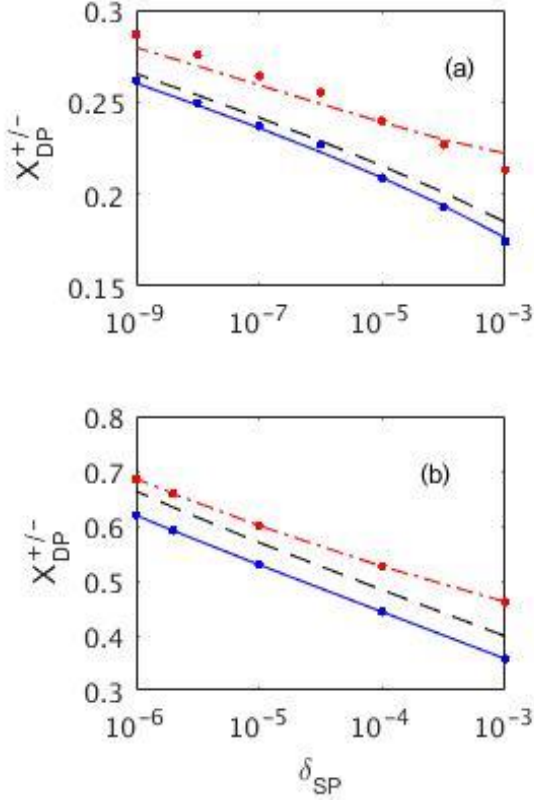

FIG. 7. Validation of the approximations (36, 39) for characteristics of the DP ( $X_{DP}^{-}$, $\bar{X}_{DP}$, $X_{DP}^{+}$ are shown by solid, dashed and dashed-dotted lines, respectively) by the simulations of the S2Z model (Eqs. (23, 24), symbols) with assigned $p$-values predicted by approximations. $\bar{i}$ =0.02 (a) and 0.2 (b).

### *2. Effect of non-uniform parameters*

After analyzing the behavior of a system with uniform parameters, subject to a certain separation at the spinodal point initially, we turn to study the system behavior when the parameters vary somewhat in space either as a step (for the S2Z model) or as an harmonic function or as spatially-random dependence (for the SD model). In this case the simulations of both models start with homogeneous IC at a certain small initial $\bar{X}$ value ($\tau = 0$: $X = \varepsilon \ll X_{SP1}$).

The simulations of the S2Z system with pre-assigned $p$-value (not presented) show a good agreement with the approximations within a wide range of the imposed non-uniformity $\alpha = 10^{-12} - 10^{-2}$ and current $\bar{i}$ =0.01-0.2.

The simulations of the SD model were conducted under imposed spatial constrains (referred to here as noise). Noise can be characterized by various parameters (amplitude, spectrum, spatial structure etc). To this end we simulate the system subject to fixed perturbations either in the form of a regular wave, $\alpha = \alpha_0 \cos(n\pi z / L_{cat})$, n=1,2,…; or subject to a white noise perturbation. We use the mean square magnitude of the perturbation as the reference parameter ($\alpha_{RMS} = \sqrt{2}/2 \alpha_0$ for a harmonic form).

A typical $X_s$ patterns simulated with imposed harmonic spatial perturbations reveal the emerging stair-case structures with order that follows the imposed perturbations: with symmetric perturbations the lithiation is initiated at both ends of the cathode i.e., at the wall and at the separator boundary (Fig. 8, n=2); with asymmetric perturbations the lithiation is initiated where $\alpha$ is largest; with multi-wave perturbations the zones of low and high $X_s$ concentrations are alternated following the profile $\alpha(z)$. Recall that at this stage of modeling the various groups are interacting only by the galvanostatic condition and not by liquid-phase interaction.

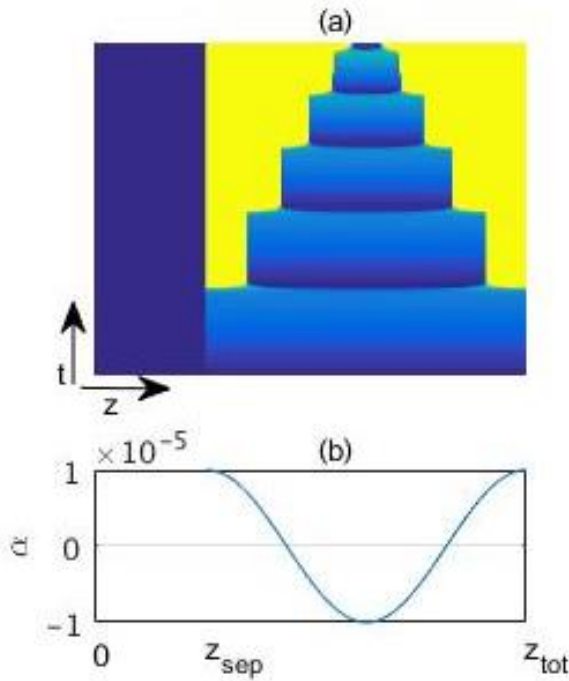

FIG. 8. The effect of non-uniformity of the harmonic form parameter perturbations on the solid concentration dynamics (a) and the imposed perturbation profile (b). Simulations with the SD model. $\alpha_0 = 10^{-5}$, $n=2$; $\bar{i} = 0.02$.

The temporal system behavior is only marginally affected by the specific profile of the imposed perturbations (compare the results with n=1, with that of n=16 shown in Fig. 9 (a)). The amplitude of the perturbation is of crucial role: With very small $\alpha_0$ the system exhibits a quasi-homogeneous behavior: symmetry breaking is not detected, the solid potential follows the equilibrium potential (shifted by the overpotential) within the whole range of $\bar{X}$ including the unstable branch. The bifurcation(s) is observed with $\alpha_0$ above a certain threshold value (actually this depends on the accuracy of calculations) at $\bar{X}_{DP} > X_{SP1}$ and is accompanied by large amplitude $\phi_s$-oscillations.

Increasing $\alpha_0$ further we observed the following tendencies: i) $\bar{X}_{DP}$ declines toward the first spinodal point, ii) the number of sequential SB events and the phase separation fraction $p$ of each break increases, and iii) the amplitude of potential oscillations declines. This tendency is elucidated by Fig. 9 (b) showing the effect of the amplitude ($\alpha_0$) on the solid potential for n=1.

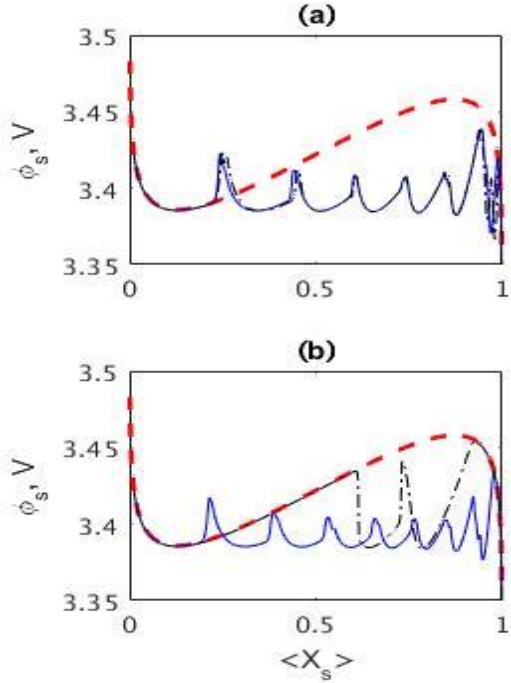

FIG. 9. The effect of the imposed harmonic noise parameter perturbations on the solid phase potential simulated with the SD model (a: $\alpha_0 = 10^{-4}$; n=1 (solid), 4 (dashed-dotted); b: n=1, $\alpha_0 = 10^{-2}$ (solid) and $\alpha_0 = 10^{-15}$ (dashed-dotted) lines). Dashed lines mark the equilibrium potential. $\bar{i} = 0.02$.

Typical $X_s$ patterns simulated under imposed white noise perturbations (fixed in time) reveal emerging structures with randomly alternated zones of low and high Li concentration (Fig. 10). With a small amplitude ($\alpha_{RMS}$) it is still possible to distinguish the sequence of the SB steps at $t = t_1, t_2, \ldots$ accompanied by potential oscillations (Fig. 11). The parameter $p$ was defined using the counters of the upper/lower Li state. Such an approach can be applied at relatively small $\alpha$ (for first SBs it is close to constant) but fails upon increasing $\alpha$ that leads to completely irregular structures.

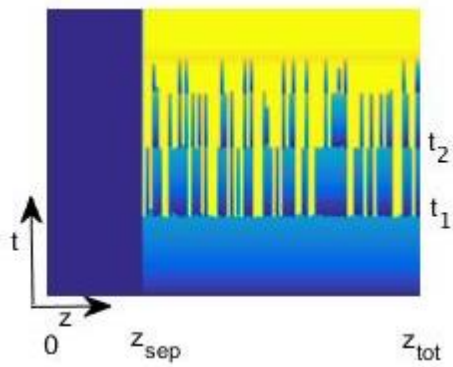

FIG. 10. The effect of the imposed white noise permanent parameter perturbation on the solid phase concentration dynamics simulated with the SD model. $\bar{i} = 0.02$, $\alpha_{RMS} = 10^{-12}$.

The effect of the magnitude of perturbation is quite similar to that in the case of the regular harmonic perturbations (compare Fig. 9 and Fig. 11). Note, that increasing noise levels leads to decreasing amplitude of the potential oscillations around the first SP.

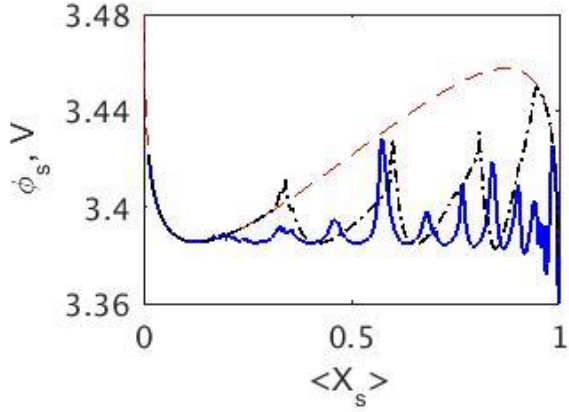

FIG. 11. The effect of the imposed white noise permanent parameter perturbations on the solid phase potential simulated with the SD model. $\alpha_{RMS} = 10^{-12}$ (dashed-dotted) and 0.1 (solid) lines, respectively; dashed lines mark the equilibrium potential $\bar{i} = 0.02$.

The approximations (36), (43) derived above are in a reasonable agreement with the simulation results: Fig. 12 plots $\bar{X}_{DP}$ and the phase separation fraction ($p$) vs the noise level ($\alpha_{RMS}$) using various profiles of the imposed perturbations. The results (both harmonic and white noise) fall into the same line. Obviously, for strong noise the DP occurs at the first SP: $\bar{X}_{DP} \simeq \bar{X}_{SP1}$, and thus one would expect an asymptotic value at high noise level. We suspect that most experimental studies belong to this asymptote. In the opposite direction ($\alpha_{RMS} \to 0$) we expect to find a homogeneous solution (predictions with $\bar{X}_{DP} > \bar{X}_{SP2}$, i.e. outside the instability domain, yield unphysical values).

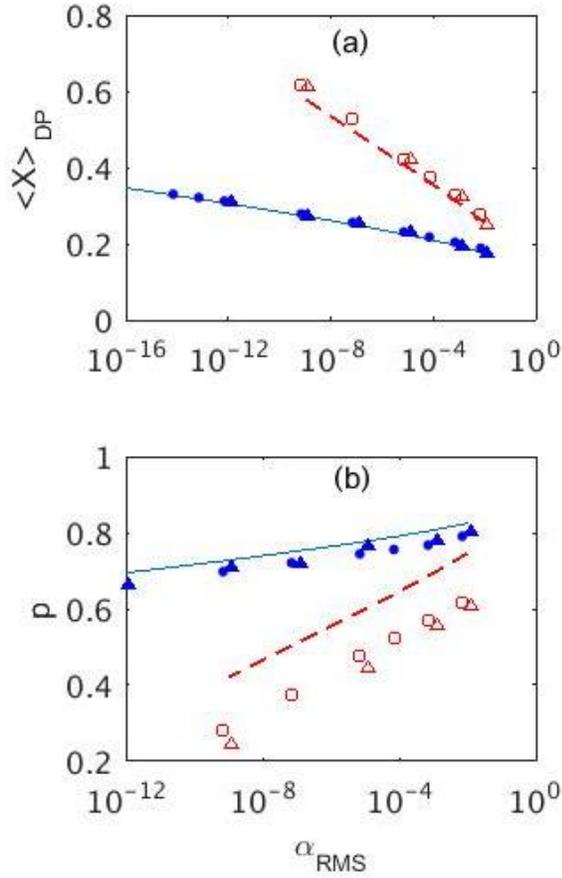

FIG. 12. Validation of the approximations (36, 43) predicting the $\bar{X}_{DP}$(a) and $p$ (b) dependence on the imposed inhomogeneity (lines) using simulations of the SD model (symbols). $\bar{i}$ =0.02 – solid lines and filled symbols; $\bar{i}$ =0.1 – dashed lines, and open symbols; points and triangles mark n=1-wave and white noise perturbations, respectively.

*Sequential SB.* After each Symmetry Breaking the domain occupied by the poor Li phase shrinks, the values of the potential peaks and the durations of steps vary (Figs. 8-11). The SB takes place when the average concentration within the poor Li domain reaches the DP value ( $\bar{X}_{DP}$ ). The comparison for sequential SB simulated by the SD model with $\bar{i}$ =0.02 and an imposed harmonic perturbation $\alpha = 0.01\cos(\pi z / L_{cat})$ for the first four cycles is shown in Fig. 13. An inspection of the results reveal a good agreement with the approximation (43-45) for the first SB, while the results diverge for the subsequent breaks. Recall, that to derive the approximation (section III) we assumed that section (iii) of the trajectory is fast, i.e. the values $\bar{X}_{DP}$ coincide with the value at the end of the separation $\bar{X}_{DP}$ (35). In the simulations we defined the DP and the EP (not shown) as the points where the low Li concentration exhibits the maximal and the minimal values, respectively. The simulations conducted with both the S2Z model (Fig. 6) and the Solid Distributed model show that duration of section (iii) is not negligible and increases with the current. Moreover, the end value $\left(X^-\right)_{EP}$ increases with current as well, i.e. an assumption on $\left(X^-\right)_{EP} \simeq 0$ yielding (36) is not valid. Thus, we expect that for the 2$^{nd}$, 3$^{rd}$ and higher SB events, the agreement between the predictions and simulations will become poorer as is shown in Fig. 13 (Still, the divergence in Fig.13 is reasonable, less than 13.% for 4th break).

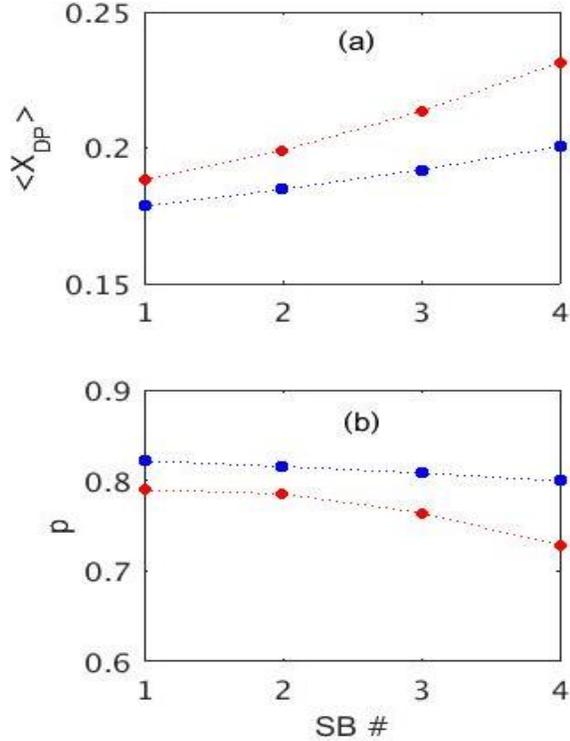

FIG. 13. Validation of the sequential SB approximations (43-45) showing the parameters at the DP as the functions of the SB event numbers (dots – simulations with the SD model; rectangulars- approximations (43-45)). $\bar{i}$ =0.02. The lines are aimed to guide the eye.

## IV. FULL MODEL ANALYSIS

In the models employed in section III above the interaction between any two points is only due to the galvanostatic condition and the spatial coordinate appears only for the convenience of presenting the system: i.e., reshuffling the various points (with the corresponding perturbation) will lead to the same results. The liquid concentration and potential ($X_l, \phi_l$) balances (Eqs. (6), (10)) induce some order in the system, since the gradients in $\phi_l$ (see Fig 2 (b), solid line at SP) are likely to make the particle near the separator to be the first to undergo SB and induce this process in the whole space (compare Figs. 1-3). We ran several simulations of the full model (still with $\sinh(\eta) \sim \eta$) with imposed perturbations.

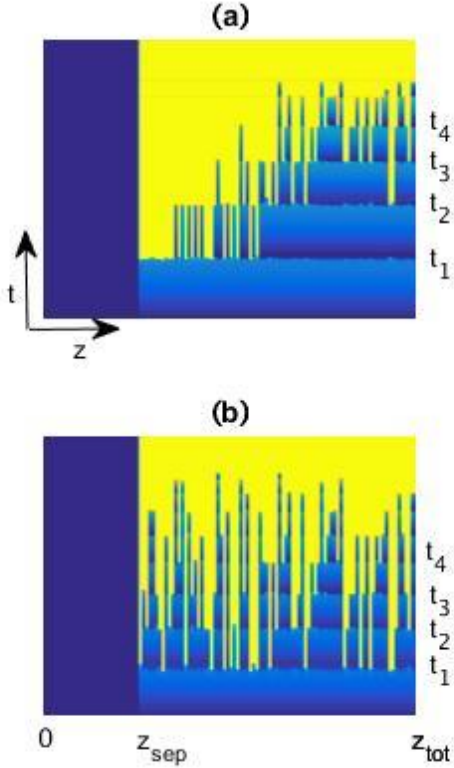

FIG. 14. The effect of the imposed white noise parameter perturbations on the solid phase concentration dynamics simulated with the full model (14)-(16). $\alpha_{RMS} = 10^{-6}$ (a) and $10^{-2}$ (b); $\bar{i}$ =0.02.

Figure 14 illustrates the gradual break up of a regular staircase structure (compare Fig. 1 without the noise) due to the imposed noise parameter perturbations. The simulated dynamics (Fig. 14) as well as the SB parameters (Fig. 15) resemble the combined effect of the liquid potential and perturbations. With large $\alpha$ the effect of perturbations dominates: the DP is close to the SP1, the functions $\bar{X}_{DP}(\alpha)$ and $p(\alpha)$ are in a reasonable agreement with approximations (43). With decreasing $\alpha$ the effect of noise decreases while the role of the liquid potential which is the intrinsic state variable is enhanced. With $\alpha$ below $\sim 10^{-6}$ (for the set of parameters in Fig. 15) the curves $\bar{X}_{DP}$ and $p$ as the function of the perturbation exhibit a plateau asymptote (dotted lines in Fig. 15) showing that the system behavior is dictated by the liquid potential.

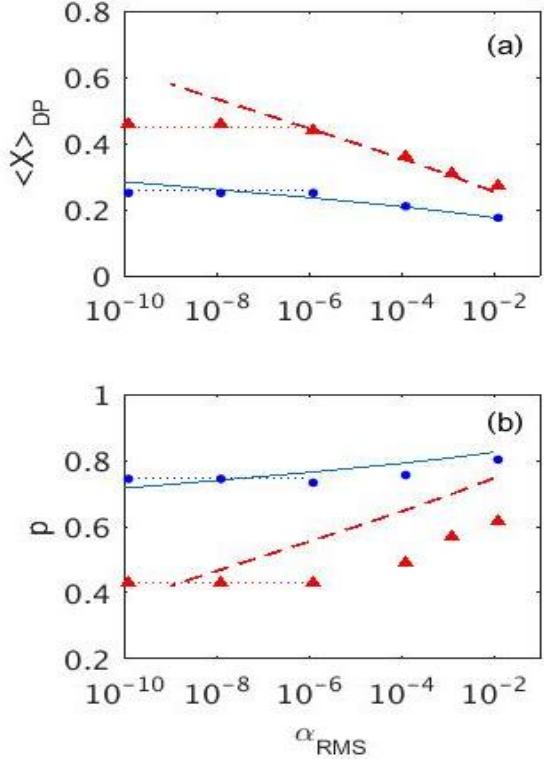

FIG. 15. The effect of the parameter perturbations on the SB characteristics (a - the solid phase concentration ($\bar{X}_{DP}$), b - the fraction occupied by the poor-charge Li, ($p$)) simulated with the full model (14)-(16) shown by symbols vs. approximations (43) shown by lines. $\bar{i}$ =0.02 (solid lines and points), $\bar{i}$ =0.1 (dashed lines and triangles). Dotted blue lines mark the constant value plateau.

The full model does not admit strictly homogeneous solutions. The liquid potential gradient will cause a potential (and consequently, over-potential) gradient. As a result there is always a 'noisy' term that will induce symmetry breaking. To quantify this effect we employ the approximations derived for the SD model (section III) to the full model. Accounting for the liquid potential we present the over-potential $\eta$ in the following form:

$$\eta = \phi_s - V_{OC} + \mu - \phi_{liq} = \eta_{SD}(1-\beta) \tag{47}$$

where $\eta_{SD} = \phi_s - V_{OC} + \mu$, is the over-potential used in the SD model and the new parameter $\beta = \phi_{liq}/\eta_{SD}$ expresses the effect of $\phi_{liq}$. Note that both $\eta_{SD}$ and $\beta$ are temporally and spatially dependent. The refined dynamic equation for the solid phase takes a form

$$\frac{dX}{dt} = -a(1-\beta)\eta_{SD} \qquad (48)$$

It seems reasonable to redefine the characteristic constant value parameter via the liquid potential drop over the cathode under the quasi-homogeneous lithiation

$$\beta^* = \frac{\phi_l(\zeta_{tot}) - \phi(\zeta_{sep})}{2\eta^*} \qquad (49)$$

Accounting for approximation (A17) while setting $X_l(0) = 1$ we obtain

$$\beta^* = b\frac{(d+g)}{4}(\zeta_{tot} - 1)^2 \qquad (50)$$

For the set of parameters used in the paper we obtain $\beta^* = 3.5 \times 10^{-6}$ which is in a good agreement with the simulation results (Fig. 2 (b) shows the dimensional value). It is useful to list the components of the parameter $\beta^*$ via the dimensional variables (17):

$$b(\zeta_{tot} - 1)^2 = \frac{(1-t_+)}{\varepsilon}\frac{L_{cat}^2}{D_{amb}}\frac{2a_p\hat{i}_0}{C_l^0 F};$$
$$d + g = \frac{3D_+ - D_-}{2(D_+ + D_-)} \qquad (51)$$

An inspection of relations (50), (51) shows that the parameter $\beta^*$ is proportional to the length of the cathode, the geometry parameter $a_p$, the exchange current coefficient $\hat{i}_0$ etc. and is independent on the averaged current $\bar{i}$. The latter result agrees with the simulations (Fig. 15) showing that the transition from the $\alpha$-dependent curves to the $\alpha$-independent plateau does not depend on the value of $\bar{i}$.

Turning back to the proposed approximations (43), we suggest to employ it to the full model using the modified parameter:

$$\tilde{\alpha} = \alpha + \beta^* \qquad (52)$$

This allows to show the similarity of the effect of the liquid potential on the SB process with that of the imposed (noise) perturbations of the physical properties.

It is of practical interest to draw in the parameter plane the domain of conditions that induce phase separation. To that end we plot (Fig. 16) the smallest current that avoids SB vs. $\alpha$ obtained by approximation (43) with several $\bar{X}_{DP}$ values calculated as a product $\kappa \bar{X}_{SP2}$ with $\kappa$=0.80, 0.85, 0.90 and 0.95 and fixed $p = 1 - \bar{X}_{DP}$. This curves almost overlap with $\bar{X}_{DP} < 0.90 \bar{X}_{SP2}$ (with higher $\bar{X}_{DP}$ the curves deviate). Under conditions used in the paper ($\beta^* = 3.5 \times 10^{-6}$) using approximations (43) yields the limiting value $\eta^* \simeq 0.12$ or the limiting current $\bar{i} = 0.24$ for the case of uniform parameters, which is in a good agreement with the simulations.

Using relations (43) with fixed $\bar{X}_{DP} < \bar{X}_{SP2}$ allowed to obtain the first insight into the problem, while in this limiting case $\bar{X}_{DP} \to \bar{X}_{SP2}$ it should be refined ($\mu'(X) \to 0$) to account for quadratic terms as well. Accounting for the modified parameter $\tilde{\alpha}$ (52) seems reasonable while the detailed analysis of this issue requires will be addressed in the future.

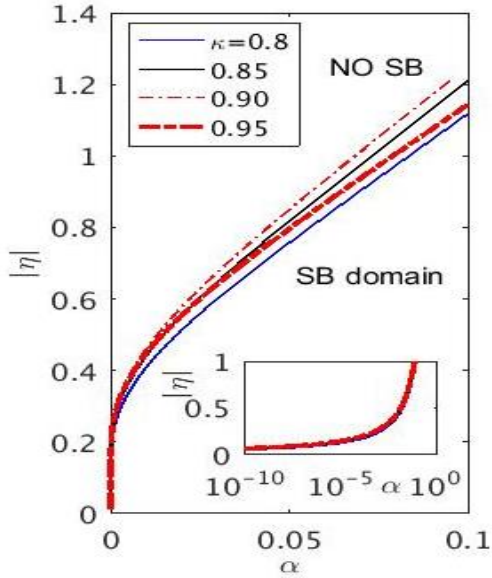

FIG. 16. The bifurcation diagram showing the required current ($\bar{i} = 2\eta^*$) as the function of the imposed perturbation ($\alpha$) with several fixed values of the solid phase concentration at the DP ( $\bar{X}_{DP} = \kappa \bar{X}_{SP2}$ ). The insert shows the plots with the logarithmic x-axis.

Concluding this section we list the main models and the corresponding results arranged in Table I.

Table I. The summary of the employed models and the corresponding the state variables, the governing equations, the specific features and the figures showing the simulation results.

| Model | Variables | Equations | Specific features | Figs. |
|---|---|---|---|---|
| Full | $X_s(t,z)$<br>$\phi_s(t)$<br>$X_l(t,z)$<br>$\phi_l(t)$ | (14)<br>(11)<br>(15)<br>(16) | Exhibits SB under small current. Liquid phase gradients affects pattern like noise effects | 1 - 4,<br>14 - 16 |
| SD (Solid Distributed) | $X_s(\tau,\zeta)$<br>$\phi_s(\tau)$ | (14)<br>(11) | Main properties like of the full model, but with infinite patterns; was used to study noise effects in IC or in spatial nonuniformity | 8 - 13 |
| S2Z (Solid 2 Zone) | $X^\pm(t)$<br>$\phi_s(t)$ | (23)<br>(24) | Exhibits SB and allows to derive analytical solutions for SB parameters | 5 - 7 |

## V. DISCUSSION

Symmetry Breaking is an important phenomenon in various scientific fields. It is usually analyzed as a bifurcation, upon changing a parameter, from one type of patterned state to another. The SB events of interest here are predicted for a batch system where the patterned state changes over very short intervals as the battery is charged or discharged.

While the phenomena of group-by-group charging (or discharging) has been claimed to occur in batteries with phase-separation thermodynamics, subject to galvanostatic control, both by simulations [10] and later by experiments [12], the origin of this phenomena was not lucid. The present work is the first to quantitatively predict this behavior and to describe its characteristics.

The full model accounts for four variables. To obtain analytical approximations we reduced it to the *solid* phase *distributed* (SD) model governed by $X_s = X_s(t,\xi)$. The latter was reduced further to a *solid two-zone* (S2Z) model using the averaged levels $X^\pm = X^\pm(t)$ of the poor/rich phases as the state variables (see Table 1 for list). All models were analyzed subject to the galvanostatic-control condition and spatially nonuniform properties (referred to as noise effects).

The SD and S2Z models admit homogeneous solutions: The symmetry may break down if initial conditions are inhomogeneous or if parameters are not spatially uniform. Both imposed perturbations were considered here. The results are highly sensitive to the imposed conditions. This issue was overlooked in previous works. Numerical results of the full model exhibit the Symmetry Break since the liquid phase potential gradient ($\phi_l$) over the cathode works like a perturbation that triggers the SB and it defines the spatial order (i.e., pattern) of SBs. In a real system the lithiation rate will depend on physical properties like activity, geometry (area to volume ratio denoted by $a_p$) etc., so that even with homogeneous initial conditions a separation will be induced by the difference in the growth rate due to nonuniform parameters.

To gain analytical results we analyzed the trajectories of the full and reduced models near the SB event as a three-stage scheme. Since the SB transition is supercritical, and its effect is initially very small, we identify the bifurcation by the local maximum of the averaged poor-Li concentration charge ($X^-$), which is clearly identifiable; the point is referred to as the Detectable Point (DP).

Using such a scheme applied to a reduced S2Z model allowed to obtain the approximate relations for the characteristics of the DP (the average solid phase concentration ($\bar{X}_{DP}$) and the fraction of the poor Li phase ($p$) as the functions of the initial perturbation (39) or as the function of the nonuniform parameter distribution (43). We detected the similarity between the liquid phase potential effect and the noise factor and defined the parameter $\beta^*$ (49) that allows to translate the obtained approximations to the full model analysis. This methodology was verified by numerical simulations.

Now we would like to comment on the main assumptions made to obtain the proposed approximations and on the possible implication or extension:

i) We employed the set of parameters that allows to simplify the reaction rate by approximation $\sinh(\eta) \simeq \eta$. Using exact relations led to similar relations.

ii) In the present paper a constant exchange current ($\Psi(C_l, C_s) = 1$ in Eq. (5)) is addressed yielding symmetric lithiation/delithiation process. Accounting for variable $\Psi(C_l, C_s)$ requires additional efforts to formulate the S2Z model and to find its analytical solution. This issue may be addressed in the following study. However since $\Psi(C_l, C_s) > 0$, we do not expect it to change the qualitative behavior.

iii) The kinetics effects of mass transfer from the liquid to solid phase were not accounted for here, and will be addressed in the future work.

iv) To obtain approximations using S2Z model several assumptions were employed:
- the last stage (iii) of the proposed trajectory around the SB event was assumed instantaneous while the low and the high values of Li concentration reach the limiting values (of 0 and 1, respectively). These approximations are valid with low current and become poor at higher current. This explains the divergence between the approximation and simulation results with increasing SB event.
- the linear approximation of the chemical potential around the average value is not valid around the spinodal points (26) but can be improved by accounting for the quadratic term.

v) In the present paper we also did not consider the temporal profile of the solid phase potential ($\phi_s(\bar{X})$); the resulting oscillations are the key evidence to SB in either numerical or experimental studies. The deviation of $\phi_s(\bar{X})$ from the homogeneous solution trajectory following the proposed scheme within the first two sections is $O(\Delta X^2)$, the extended analysis will be addressed in the future work.

## VI CONCLUSIONS

The present work describes the SB events in a transient system, rather than in an open continuous system. The SB event was claimed to be the mechanism that qualitatively explains the morphogenesis starting with Turing [1], that employed a continuous system to demonstrate it, and there are several systems that verified the conjecture that physiological patterns emerge by SB. The origin of morphogenesis has and is attracting considerable attention by Developmental Biologists and by theoreticians. SB events in a transient (batch) system may be more adequate to account for morphogenesis.

The present work is the first to predict the main characteristics of the sequential group-by-group charging such as the average Li concentration and the fraction of the low Li part at the SB points. We propose the governing parameter $\tilde{\alpha}$ (52) that accounts for the complex effect of the liquid potential drop over cathode and the noise factor. A bifurcation diagram is constructed showing the low current boundary that ensures a homogeneous lithiation/delithiation as the function of this parameter.

To decide whether the transition occurs as group by group or particle by particle other effects like thermal effects and solid diffusion within the particle, should be accounted for. The transition was predicted to occur in order of increasing particle size[30]. Thesse questions are of important also to ther nonlinear miscrostructured electrochemical systems like CO electrooxidation on Pt [31].

**Abreviatures**

DP detectable point

S2Z solid two-zone (model)

SB symmetry break

SD solid distributive (model)

SP spinodal point

# APPENDIX A: APPROXIMATION OF THE SS LIQUID CONCENTRATION AND POTENTIAL PROFILES UNDER THE HOMOGENEOUS LITHIATION.

We choose the length scale $L_0 = L_{sep}$ to simplify the notation.

*The liquid phase balance* (15) under SS conditions can be presented in the following form

$$\frac{\partial^2 X_l}{\partial \zeta^2} = \begin{cases} 0 & 0 < \zeta < 1 \\ s & 1 < \zeta < \zeta_{tot} \end{cases} \quad (A1)$$

where

$$s = b \sinh(-\eta^*) \quad (A2)$$

subject to BC

$$\zeta = 0: \left.\frac{\partial X_l}{\partial \zeta}\right|_0 = -s(\zeta_{tot} - 1) = -S; \quad \zeta = \zeta_{tot}: \frac{\partial X_l}{\partial \zeta} = 0; \quad (A3)$$

Solution of (A1) presents a linear function $X_l(\zeta)$ within the separator and a parabolic one within the cathode. Using a matching conditions (at $\zeta = 1$) we obtain

$$X_l = \begin{cases} X_l(0) - S\zeta, & \zeta < 1 \\ X_l(0) - s(\zeta_{tot} - 1) + \frac{s}{2}\left[(\zeta - 1)(1 + \zeta - 2\zeta_{tot})\right] & 1 < \zeta < \zeta_{tot} \end{cases} \quad (A4)$$

yielding

$$\frac{\partial X_l}{\partial \zeta} = \begin{cases} -s(\zeta_{tot} - 1) & 0 < \zeta < 1 \\ -s(\zeta_{tot} - \zeta) & 1 < \zeta < \zeta_{tot} \end{cases} \quad (A5)$$

*The liquid potential* profile is obtained using Eq. (16) that accounting for (A2) takes a form

$$\frac{\partial}{\partial \zeta}\left(X_l \frac{\partial \phi_l}{\partial \zeta}\right) = -ds - g\frac{\partial^2 X_l}{\partial \zeta^2} \qquad (A6)$$

subject to BC

$$\zeta = 0: \phi_l = 0; \quad \zeta = \zeta_{tot}: \frac{\partial \phi_l}{\partial \zeta} = 0 \qquad (A7)$$

*Separator* domain. Within a separator Eq. (A6) accounting for a linear profile $X_l(\zeta)$ Eq (A4) yields

$$X_l \frac{\partial \phi_l}{\partial \zeta} = const = X_l \frac{\partial \phi_l}{\partial \zeta}\bigg|_0 \qquad (A8)$$

To obtain this constant we integrate Eq. (A6) over the total domain:

$$X_l \frac{\partial \phi_l}{\partial \zeta}\bigg|_{\zeta_{tot}} - X_l \frac{\partial \phi_l}{\partial \zeta}\bigg|_0 = -sd(\zeta_{tot} - 1) - g\left[\frac{\partial X_l}{\partial \zeta}\bigg|_{\zeta_{tot}} - \frac{\partial X_l}{\partial \zeta}\bigg|_0\right] \qquad (A9)$$

or accounting for the BC (A3), (A7) at $\zeta = \zeta_{tot}$

$$-X_l \frac{\partial \phi_l}{\partial \zeta}\bigg|_0 = -Sd + g\frac{\partial X_l}{\partial \zeta}\bigg|_0$$

Accounting for the BC (A3) with $\zeta = 0$ the latter is reduced to following

$$-X_l \frac{\partial \overline{\phi}_l}{\partial \zeta}\bigg|_0 = -S(d+g) \qquad (A10)$$

Now Eq. (A8) reads

$$X_l \frac{\partial \phi_l}{\partial \zeta} = const = S(d+g) \qquad (A11)$$

Accounting for profile (A4) we obtain

$$\frac{\partial \phi_l}{\partial \zeta} = \frac{S(d+g)}{X_l(0) - S\zeta} = \frac{S(d+g)}{X_l(0)} \frac{1}{1 - S/X_l(0)\zeta}$$

yielding

$$\phi_l(\zeta) = (d+g)\log\frac{X_l(0) + S\zeta}{X_l(0)}$$

that can be reduced ($S/X_l(0) \ll 1$) to the linear relation:

$$\phi_l(\zeta) = (d+g)\frac{S}{X_l(0)}\zeta \qquad 0 < \zeta < 1 \tag{A12}$$

*Cathode domain* Integration of (A6) over the cathode domain ($1 < \zeta < \zeta_{tot}$) results:

$$X_l\frac{\partial \phi_l}{\partial \zeta} - X_l\frac{\partial \phi_l}{\partial \zeta}\bigg|_1 = -(\zeta - 1)sd - g\left[\frac{\partial X_l}{\partial \zeta}\bigg|_\zeta - \frac{\partial X_l}{\partial \zeta}\bigg|_1\right] \tag{A13}$$

Accounting for (A5) and (A11) applied with $\zeta = 1$ and certain rearrangements we obtain

$$X_l\frac{\partial \phi_l}{\partial \zeta} = (d+g)s(\zeta_{tot} - \zeta) \tag{A14}$$

To obtain profile $\phi_l(\zeta)$ Eq. (A14) should be integrated accounting for $X_l(\zeta)$ defined by Eq. (A4). Accounting for marginal deviations $X_l(\zeta)$ (see Fig. 2) to the first approximation we set $X_l(\zeta) = const$ and obtain the potential drop over the cathode as following

$$\phi_l(\zeta) - \phi_l(1) = \frac{(d+g)s}{2X_l(0)}\left[(\zeta_{tot} - 1)^2 - (\zeta_{tot} - \zeta)^2\right] \quad 1 < \zeta < \zeta_{tot} \tag{A15}$$

We are interesting on the potential drop over the cathode:

$$\phi_l(\zeta_{tot}) - \phi_l(1) = \frac{(d+g)s}{2X_l(0)}(\zeta_{tot} - 1)^2 \tag{A16}$$

Finally accounting for (A12) we obtain

$$\phi_l = \begin{cases} (d+g)\dfrac{s(\zeta_{tot}-1)}{X_l(0)}\zeta & \zeta < 1 \\ \dfrac{(d+g)s}{2X_l(0)}\left[(\zeta_{tot}-1)^2 - (\zeta_{tot}-\zeta)^2\right] + \phi_l(1) & 1 < \zeta < \zeta_{tot} \end{cases} \tag{A17}$$

## APPENDIX B: APPROXIMATIONS OF THE DP CHARACTERISTICS FOR THE NON-UNIFORM PARAMETER DISTRIBUTION.

*Dynamics equations* (39) within trajectory sections (i) & (ii) accounting for approximation (26) can be presented as following:

$$\begin{aligned}\frac{dX^-}{dt} &= -a\left(1 - \frac{\alpha}{p}\right)\left(\eta^* + \mu'_{\bar{X}}(X^- - \bar{X})\right) \\ \frac{dX^+}{dt} &= -a\left(1 + \frac{\alpha}{1-p}\right)\left(\eta^* + \mu'_{\bar{X}}(X^+ - \bar{X})\right)\end{aligned} \tag{B1}$$

System (B1) while neglecting the term of $\sim \alpha(X^+ - X^-)$ can be reduced to the following nonhomogeneous equation

$$\frac{d(X^+ - X^-)}{dt} = -a\left[\mu'_{\bar{X}}(X^+ - X^-) + \eta^* \frac{\alpha}{p(1-p)}\right]$$

Or, accounting for Eq. (14), the latter can be presented as

$$\frac{dy}{d\bar{X}} = \frac{\mu'_{\bar{X}}}{\eta^*} y + \frac{\alpha}{p(1-p)}, \quad y = X^+ - X^- \tag{B2}$$

Eq. (B2) differs from Eq. (28) derived for a uniform properties by the last (non-homogeneous) term.

We search the solution of (B2) as the solution of the homogeneous part (28) with a variable pre-exponent factor:

$$y_{partic} = D(\bar{X}) \exp\left(\frac{\mu(\bar{X})}{\eta^*}\right) \tag{B3}$$

yielding

$$y'_{partic} = \left(D\frac{\mu'_{\bar{X}}}{\eta^*} + D'\right) \exp\left(\frac{\mu(\bar{X})}{\eta^*}\right) \tag{B4}$$

Substiting (B3) and (B4) in (B2) we obtain

$$\left(D\frac{\mu'_{\bar{X}}}{\eta^*} + D'\right) \exp\left(\frac{\mu(\bar{X})}{\eta^*}\right) = \frac{\mu'_{\bar{X}}}{\eta^*} D \exp\left(\frac{\mu(\bar{X})}{\eta^*}\right) + \frac{\alpha}{p(1-p)} \tag{B5}$$

which is reduced to an equation with respect to $D(\bar{X})$:

$$D'\exp\left(\frac{\mu}{\eta^*}\right) = +\frac{\alpha}{p(1-p)}$$

yielding

$$D(\bar{X}) = D_0 + \frac{\alpha}{p(1-p)} \int_{\bar{X}_{IC}}^{\bar{X}} \exp\left(-\frac{\mu(\bar{X})}{\eta^*}\right) d\bar{X} \tag{B6}$$

The total solution of (B2) follows

$$y(\bar{X}) = \left[D_0 + \frac{\alpha}{p(1-p)} \int_{\bar{X}_{IC}}^{\bar{X}} \exp\left(-\frac{\mu(\bar{X})}{\eta^*}\right) d\bar{X}\right] \exp\left(\frac{\mu(\bar{X})}{\eta^*}\right) \tag{B7}$$

Assuming that the initial deviation is null ($t=0$: $\bar{X} = \bar{X}_{IC}$, $y=0$) we obtain (B8)

$$y = \frac{\alpha}{p(1-p)} \exp\left(\frac{\mu}{\eta^*}\right) \int_{\bar{X}_{IC}}^{\bar{X}} \exp\left(-\frac{\mu(\bar{X})}{\eta^*}\right) d\bar{X}$$

or

$$X^+ - X^- = \frac{\alpha}{p(1-p)} J(\bar{X}) \exp\left(\frac{\mu}{\eta^*}\right); \tag{B9}$$

where

$$J(\bar{X}) = \int_0^{\bar{X}} \exp\left(-\frac{\mu(\bar{X})}{\eta^*}\right) d\bar{X} \tag{B10}$$

*Current balance equations at the DP.* Similarly to the case of constant properties (Eq. 31) we obtain at the DP the following relations:

$$\eta^- = \phi_{sDP} - V_{OC} + \mu_{sDP}^- = 0;$$
$$\eta^+ = \phi_{sDP} - V_{OC} + \mu_{sDP}^+ = \eta^* / \left[\left(1 + \frac{\alpha}{1-p}\right)(1-p)\right] \tag{B11}$$

System (B11) can be rearranged yielding

$$\mu_{DP}^+ - \mu_{DP}^- = \frac{\eta^*}{(1+\alpha/(1-p))(1-p)} \tag{B12}$$

Eq. (B12) accounting for approximation (26) can be presented as following

$$\mu'_{\bar{X}}\left(X^+ - X^-\right) = \frac{\eta^*}{(1+\alpha/(1-p))(1-p)} \tag{B13}$$

System (B9), (B13) allows to obtain

$$\frac{\eta^*}{\mu'_{\bar{X}DP}(1+\alpha/(1-p))} = \frac{\alpha}{p} J(\bar{X}_{DP}) \exp\left(\frac{\mu(\bar{X}_{DP})}{\eta^*}\right) \tag{B14}$$